# arXiv and the Symbiosis of Physics Preprints and Journal Review Articles

Brian Simboli, Ph.D.[1]

## Abstract

As the current regime of scholarly publishing changes in the face of high pricing and the unsustainably large burden it imposes on peer reviewers, new thinking will need to emerge about how to reform publishing along lines that best meet two perennial needs of scientific communication. One is the need to claim priority of discovery. The second is to provide integrative review of research programs.

This paper discusses a "model" that addresses these two needs with respect to one subject area for which, currently, it is most likely suitable: physics.

The paper makes no assumptions that this model will ever be fully realized. So why even propose it? It aspires merely to be a heuristic or guidepost, in three ways: it provides an analytical framework for criticizing aspects of the current publishing ecosystem; it helps diagnose problems in current efforts to reform it, including those emanating from the open access movement; and it raises consciousness about certain emphases that could gradually enrich scholarly publishing.

The paper argues that preprints, whose use is increasingly important in science and already well-established in physics, are properly the vehicle for claiming priority of discovery and for eliciting feedback that will help with draft versioning. Traditional journal publishing in physics, however, should much more focus on providing synthesis in the form of overlay journals that play the same role as review articles. The goal is a much greater symbiosis between these two emphases.

While in its ideal form, physics communications would consist of review articles that synthesize and critically evaluate research trends addressed in preprints, the paper suggests two independent emphases that can imperfectly realize the model's ideals for physics communication: greater reliance on review articles, and increased citing of preprints within traditional journal publishing.

In developing its theses, the paper locates physics preprints—with a focus on arXiv--within the history of scientific communication, then outlines the model and discusses how it addresses problematic aspects of the traditional model of peer-reviewed journal publishing. Proposed enhancements to arXiv can help incrementally to advance the model's goals. Finally, librarians can bring their special expertise to bear in helping to advance the model.

---

[1] The views in this essay are not those of the author's employer. Author's email: brs4@lehigh.edu

Given the widespread adoption of preprints among physicists (not to mention other highly quantitative fields), and given that the model places a great deal of emphasis on preprints, the discussion below focuses on physics. But many of the problems with physics publishing also beset other fields of science, and there is evidence of increased reliance on preprints outside physics. Therefore, some of the points below may prove applicable in other areas of science.

Given the length of this paper, readers should use the table of contents as a guide to help identify portions of the analysis that might be of interest to them.

## Table of contents

7. Commenting and annotation features
8. Granular classification or thesaurus schemes/abstracting and indexing
9. Manuscript template and uniformity of manuscript style
10. Labeling of review articles
11. Moderated submission review, slightly expanded
12. Manuscript submission capability

- Relevance of the model for non-physics subjects
- Preprints and Science Librarianship
- Conclusion
- Acknowledgments
- Bibliography

**Introduction: A Critique of the Open Access Movement**

The millennium's turn saw a huge flurry of impassioned listserv discussion about new models of journal publishing. Much of this discussion lamented the high and increasing costs of journal subscriptions and called for open access (OA) to articles.

The post-history of those innovative discussions dashes any hopes of dramatic, near-term changes that will address problematic aspects of traditional journal publishing (hereafter, TJP), i.e., academic journal publishing that typically relies on peer review. This is so for various reasons explored later, which have to do with the fact that in the last twenty years, no stakeholder within the journal-publishing ecosystem is likely to emerge to effect needed reform.

For now, it is important for what follows to provide a critique of how one such stakeholder, the library profession (considered in the aggregate, despite a minority that agrees with this paper's analysis or parts of it), has through its advocacy of the open access movement diverted attention from the complexities of the economics of publishing.

In brief, often the link between the two goals of OA and of reducing prices (or year over year price increases) has been unclear. The impetus to make a large quantity of articles OA has made great strides, but the price of publishing has as a result redistributed from libraries toward other funding sources, or involved burdening already stressed library budgets through library support of author publishing charges for OA. Specifically, OA advocates have not always made clear their views on how OA journal publishing does not just replicate troublesome aspects of the economics of toll access publishing. (For some particular comments about the pricing issue in relation to OA, see below.) The focus on providing OA has also distracted from another problematic aspect of scholarly publishing, the ever-increasing glut of journal articles.

For lower journal prices and OA to materialize concurrently, and to contract the range of journals, the scholarly publishing system will have to change in fundamental ways to preserve the two major needs of scholarly communication in the sciences. Given that an agent or agents of reform are unlikely to materialize, the only hope is that journal markets will sort out the problems inherent in scientific communication—via a kind of Hayekian spontaneous order-- in ways that meet these two needs.

The model proposed below can help provide a framework for addressing this problematic. The focus will be on physics publishing, for which the model is currently more suitable than in other areas of science publishing. This is because the model builds on the huge quantity of OA material ***already long available*** in arXiv.

It posits ideals that will probably not materialize in the near run, if ever, but at least the proposed model helps to highlight new emphases that can become more thematic in scholarly publishing.

How will the emerging importance of preprints address problems inherent to TJP? The definition of "preprint" in this paper will follow (Velterop 2018, p. 1) as "publication before peer review has taken place." A related but distinct question concerns what influence they *should* have in addressing those problems. The answer to the first is anyone's guess. Rather than try to predict the future, the essay merely argues that preprints have considerable potential to transform the face of TJP publishing, despite the many barriers to this happening.

Given recent indications of increasing interest in preprints and emerging challenges to TJP in the U.S. and Europe, now is time for a measured assessment of the role that the burgeoning reliance on preprints can *potentially* play in reforming scholarly publishing in physics.

Anyone familiar with the preprint landscape knows that a host of new preprint servers have now appeared with the "--xiv" suffix. The verdict is still out on what foothold these new preprint servers will establish in the disciplines outside those covered in arXiv. They are certainly making inroads as an acceptable publication format in biology. But the focus of this paper is, again, physics.

**Brief synopsis: four main theses**

Here are the four main theses that the paper defends.

- As already mentioned, the open access movement has failed thus far to address problems (notably, increasingly high pricing in excess of the CPI, as well as the problem of publishing "glut") in a systematic way. Some current initiatives in attempts to reform scholarly publishing actually perpetuate and exacerbate aspects of the problem.
- TJP has serious flaws that the publishing model discussed in the essay provides heuristics for reform, at least in the arena of physics. It is an ideal type that may never be realized, but even small, incremental efforts toward realizing it are valuable to realize the two perennial needs of scientific communication.
- The arXiv enhancements proposed in the essay, if funding were available, could help incrementally realize the potential of preprints by giving physics preprint publishing more of the look of TJP.
- The library profession and its consortial or system-wide representatives can play a part, however realistically limited, in realizing the ideals implicit in the model.

To maintain focus and because of its longstanding role as exemplar and driver of preprint publishing, the emphasis will be on the preprint server arXiv—a huge, ***already existing*** repository of OA materials--and its potentials for transforming journal publishing. While arXiv covers a variety of quantitative subject areas, again to keep focus, the emphasis will be on TJP in physics.

In developing the four theses, the paper looks at the history of scientific communication. Preprints play the same role in some respects that scientific correspondence did in an earlier day. It then describes a publishing model specifically for physics that ascribes an enhanced role to preprints vis-à-vis TJP and a correspondingly diminished, though still significant, role for the traditional journal format. Very broadly, preprints and journals should primarily focus on disclosure of ideas and on synthesizing research disclosed in preprints, respectively.

The paper next identifies various problems with physics journal publishing (such as its glut of papers, not unique of course to physics) and discusses how the model can address them.

It then identifies barriers to preprints playing a role in challenging aspects of physics journal publishing. It goes on to discuss enhancements to arXiv that may help incrementally to overcome some of these barriers. It briefly discusses the advent and progress of a number of other preprint repositories in non-physics areas that contain the -Xiv "suffix".

At some points, the paper suggests that the model it advocates accords with core values of librarianship, which can play a major role in promoting the publishing model it advocates.

The essay concludes that even though some aspects of the suggested model may never materialize, at least pristinely, the arXiv enhancements suggested can help reinforce the symbiosis between TJP and preprints.

Finally, the essay makes no pretense to have achieved anything approaching exhaustive coverage of the relevant literature about the vast array of sub-topics relevant to a full exposition of the Model. That is not its point, which instead is to provide a broad framework of analysis for reforming publishing, after providing historical context and perspective plus pointers to *some* of the relevant literature. (As always, one can take the citations mentioned below and explore the citing literature as a powerful way to develop a fuller bibliography.)

**arXiv's odyssey**

ArXiv, the cross-disciplinary exemplar of preprint publishing, launched in 1991. See (Lariviere 2014, p. 1159) and (Pepe 2017, pp. 1-2) for details about its history. Prior to the wide use of arXiv, physicists distributed preprints in paper format. This essay's author dismantled a sizable preprint collection in the 1990's as electronic distribution grew in importance.

As of March 16, 2019, arXiv reports that it provides "open access to 1,512,662 e-prints in the fields of physics, mathematics, computer science, quantitative biology, quantitative finance, statistics, electrical engineering and systems science, and economics." For a breakdown by subject of number of submissions to arXiv see here.

**History of scholarly communication in physics: two perennial needs**

The history of scientific communication helps contextualize the current-day role of preprints. Preprint servers such as arXiv now play an important and highly efficient way to satisfy a perennial need for a particular type of communication, namely rapid communication that establishes priority of discovery but more importantly enables dynamic, clamorous exploration of ideas eventually published in static venues such as journals and books.

In other words, the symbiotic relationship between preprint and journal publishing defended below replicates earlier forms of symbiosis between two forms of communication. These are a dynamic one that allowed for dynamic, unsettled, *ex ante* expression of views, and a far more staid, settled, *ex post* forms of communication that summarize conclusions reached by a well-developed research agenda or sub-discipline has evolved in noteworthy ways. Starting in early modern science, if not before, both forms have played and will continue to play complementary roles.

Consider, first, the role correspondence played in early scientific communication. (Ziman 1968 cited in Lariviere, page 1158) that "it [a preprint] is a mechanized version of the decent and proper custom of writing to one's friends, colleagues and rivals about one's current work." The following passage from (Rusnock 1999, p. 156) elaborates the role of correspondence in early modern science:

> Communications from far-flung correspondents became part of the Society's practice following Henry Oldenburg's initiative in the 1660s and 1670s, the earliest years of the Royal Society. The reading and discussion of scientific letters at regular meetings constituted a significant contribution to the intellectual and social vitality of the Society. In principle, the Society maintained extensive records to document these exchanges, including minutes of the weekly meetings, letter books that summarized all letters and replies read at the Society's meetings, and finally the publication of the Philosophical Transactions. In practice, correspondence held certain advantages over other types of activity at the Royal Society. Unlike weekly meetings of the Society, correspondence allowed geographically remote individuals to engage in, and with, the new sciences. While publication and distribution of the Philosophical Transactions certainly contributed to the diffusion of knowledge, it did not provide for the flexibility, openness, manoeuvrability and relative rapidity of interaction that correspondence did. In short, the Society's correspondence encouraged a more participatory science.

This passage suggests that correspondence provided "relative rapidity of interaction" in contrast to journal publishing. Moreover, *Philosophical Transactions* "did not provide for the flexibility, openness, manoeuvrability … that correspondence did."[2] In these respects, correspondence played a similar role to preprints. After all, these are traits of preprint publishing, which enables successive drafts of results in response to criticism. There are disanalogies, but they are not important for the general point. One is the long delay in communication of ideas through correspondence versus immediate, open access dissemination of ideas via preprints. Also, correspondence involves two way interactions whereas preprints do not. Still, the latter can prompt instantaneously transmitted interactions in the form of email exchanges as well as comments posting in electronic forums (about which, below.)

---

[2] (Guedon 2001, p. 6) qualifies this positive characterization of correspondence: "…Galileo had sent an anagram of the phrase describing his discovery of Jupiter's satellites…to Kepler (and many others) in order to establish his priority. The ideas was to place a potential rival in the uncomfortable position of reluctant witness. Galileo's move was somewhat awkward, relatively idiosyncratic, certainly brilliant; it also shows how difficult it was to assert, let alone prove, something like ownership of ideas or 'intellectual property' in the early part of the 17th century. However, Galileo's move also incorporated a strong potential for divisiveness that could weaken the whole house of science. A public registry of discoveries could help steer away from such dangerous shoals…."

More importantly, the passage suggests that in those early days of scholarly publishing, correspondence coexisted with the journal format. This is true of the current day relationship between preprints and journals. The comment from above that "letter books that summarized all letters and replies read at the Society's meetings" suggests that a difference between then and now is that correspondence had a recognizably higher status than preprints have today.

In the subsequent history of scientific communication, the need to satisfy the (relatively) rapid communication that correspondence played in those early days did not go away. It just took new forms, namely preprints and letters to the editor. (Delfanti 2015, p. 2) discusses "epistolary communication" and preprints in the same vein:

> Whereas epistolary communication has always been one of the pillars of modern science, since the end of World War II particle physics has institutionalised the practice of exchanging preprints by post, irrespective of the distribution of the same papers via science journals. The libraries of departments or laboratories used to keep an archive of preprints that used to be sent in by other schools or laboratories around the world.

Letters to the editor appears to have played a role comparable to the role mailed preprints did. Like correspondence, letters printed in the journal format enabled speed of communication, even if these letters (*qua* printed) had a finalized quality, unlike the fluidity ("flexibility, openness", to borrow Rusnock's terms) that correspondence provided in an earlier day.

(Hartman 1994, pp.146) mentions that *Physical Review's* "Letters to the Editor" "was established in July 1929 in Volume 34 to facilitate and hasten the publication of work outstandingly important and urgently required." Hartman notes a submission time requirement of "three days before date of publication!" (p. 146) and that the letters section "would ultimately evolve into another separate REVIEW supplement—PHYSICAL REVIEW *Letters*." Hartman continues (pp. 162-163) that:

> *Letters to the Editor* was a too popular feature of our journal. [There was] …. a full page admonishment "To Contributors" from the Editors. Five years of the Section had seen nice growth, but now, concerns: the most serious: "…growing tendency among contributors to be satisfied with the hasty, incomplete, and often inadequate record of their investigations…" which Letters provides. Few enjoy writing up the record the report "…when the primary urge to secure priority can be satisfied by dashing off a Letter to the Editor." Further, "It was intended neither that it (Letters) be a place for the preliminary announcement of all work, nor that, in the fields covered, it should replace more formal and critical articles. If the present tendency to record much of the important work in several laboratories by a series of Letters to the Editor of gradually increasing length is continued…" standards of the PHYSICAL Review will be "seriously lowered."

The inaugural issue of *Physical Review Letters* was July 1, 1958 (see *Physical Review Letters* 1958 for editorial.) Concerning the role letters to the editor of *Nature* played, and Rutherford's role with it, see (Baldwin 2014, p. 259).

Given the resounding success of arXiv as a means of fulfilling many of the same functions that correspondence did in an earlier day, one can agree with biochemist David Green's assertion that an early, failed effort to create preprints in biology had been (Cobb 2017, p. 3, Fig. 1) "'one of the most revolutionary innovations in the history of science communication'."

This sentiment echoes the quotation from S. Gass cited in (Marra 2017, p. 373) that "the creation of ArXiv, … has been recognized as 'the most significant change in scientific communication since the establishment of the journal in the $17^{th}$ century'." This comparison to the advent[3] of the scholarly journal is not hyperbolic. Immediate electronic access to a massive, organized quantity of research disclosures—for *free* to the public—is truly a remarkable achievement in intellectual history.

In summary, physics preprints now satisfy in a very effective way the long-existing need for fluid, rapid communication in that field. The next section describes a model for a publishing future that elevates, to some extent, the status of preprints as a format of scholarly communication and demotes, to some extent, the importance of journals. It strives for greater balance between these formats and greater appreciation for their symbiosis.

**The "Model": an enhanced role for preprints in scholarly publishing; a contracted, re-purposed role for journals**

This section describes a scholarly publishing model (hereafter, the "Model"[4]), that accords an enhanced scholarly communication role to preprints and a somewhat diminished one to journals. The Model attempts to correct the unbalanced value that academia ascribes to each and address the oddity of two large repositories of research through which researchers have to wade, one consisting of preprints, the other journal articles.

This paper does not assume that Model will ever materialize in its pristine form. The Model is (to borrow a Weberian term) an "ideal type" that can guide reform of publishing in small ways. It is heuristic in three ways. First, it suggests emphases that should emerge if the two salient needs of scientific communication are to be met. Second, to the extent that any agent emerges to effect reform in scholarly publishing, it provides an ongoing regulative ideal. Third, it provides a framework to assess critically that value and success of any such reforms.

That being said, it is also essential to recognize that if the past twenty years or so indicates anything, it is highly unlikely that any agents of reform will materialize, at least in any significant sense. To assume otherwise is naive. Nonetheless, there can and should be small, incremental efforts to accelerate slowly what market dynamics of their own accord may achieve if all goes well in the long run. Will all go well in the long run? That is anyone's guess, but even small, incremental reforms over a long period of time and along the lines envisioned are more useful than the direction scholarly publishing has taken in the last twenty years (or more accurately, not aken.)

---

[3] For a bit of history, see the Lehigh University online exhibition, "The Royal Society and the Origins of Scientific Communication."

[4] Model is henceforth capitalized throughout the essay, not to reify, nor as an affectation, but as a reminder that the "model" discussed throughout the paper is the one mentioned in this section.

Axiomatic to the Model is the principle that preprints and journal articles are primarily for disclosure of ideas and for synthesizing research disclosed in preprints, respectively.

Before going on, it is worth pausing to quote comments from (Sandweiss 2009) that reflect the views of the then *Physical Review Letters* editor. Sandweiss, a Yale physics professor, brilliantly points to serious problems in science communication in physics and, indeed, in the practice of physics itself:

> The most difficult problems that face the future scientific publishing enterprise are those arising from the ever-increasing volume of published scientific research. …
>
> The individual scientific reader is then reduced to reading an ever-smaller segment of published results and to relying on meetings, conferences, selective perusal of the arXiv, and the "grapevine." That leads to an increased narrowing of the scientific background and depth of the typical scientist and of the scientific community as a whole.
>
> Such a narrowing has several unfortunate effects. The individual scientist will be more likely to miss a development in a related field (or even in his or her own field) that could be important to his or her own research. In the past, a physicist knew enough about other physics subfields that it was possible to change from one subfield to another with relative ease, … . So such transitions will be less well motivated and less likely to occur. The past is replete with major discoveries that arose, at least in part, from such "transplantations."
>
> The unity of physics and the ability of physicists to broadly understand most of physics has made physicists particularly useful in a variety of social and policy roles. That breadth physicists have had is gradually (or not so gradually) disappearing. …
>
> Of course it is natural and desirable that research continues and that its results are peer reviewed and recorded for use by others now and in the future. …. It is the scientific publishing enterprise that must work to deal with the expansion and evolution of scientific information.
>
> There are promising developments on the horizon that are aimed at the problem. The examples I mention below come from physics because whatever expertise I have lies there. Without doubt other areas of science have analogous cases.
>
> One example is the development of virtual journals, which collect links to articles in a particular field from many different journals. These journals, such as the *Virtual Journal of Quantum Information* and the *Virtual Journal of Biological Physics Research*, relieve the subscriber from having to search many different journals for articles in a given field.
>
> Another development is the new APS journal *Physics*, which is intended to be accessible to all physicists. It publishes accounts of important articles in the APS journals as well as links to those articles.

A more speculative idea is the development of a computer program that will be able to "interview" each physicist to find out what articles he or she would have selected if he or she had indeed been able to read all of the scientific literature. The interview would identify the number of articles each physicist would select and his or her personal priorities with respect to subfields. Such an artificial intelligence (AI) program would then read all the literature and select the relevant articles to provide to the subscriber.

The journals would continue the way they are now. The Editors, just as they do now, would work to maintain and improve the journals' standards. However, the journals would become the database upon which the AI program operates. I omit the interesting discussion of how the business model of the operation of such a program would work.

Clearly such an AI program is highly ambitious. However, physics does have various features that would make such a development easier than an AI program that would mirror an individual's interests on a broader scale.

Some AI efforts are already under way. For example, Google (more specifically Google Scholar) covers peer-reviewed papers, theses, abstracts, and other scholarly literature from all broad areas of research—a collection of material that might loosely be termed the "Academic Web." This collection would be part or related to parts of the AI system I have described. Such an AI program would be very helpful to the research part of the narrowness problem but would probably not do much for the unity of physics. To maintain the breadth of physicists, projects like *Physics* or various evolutions of that journal are needed.

It seems very likely that the future of scientific publishing will involve major innovation in electronic aids to read the literature and in new journals with broad goals that wrestle with the flood of information.

The Model discussed below, *if* ever fully implemented, would address these problems. (Again, the necessary refrain, lest the reader miss its point: no assumption is made here that it will ever be achieved in pristine form.) In particular, its proposed new direction for journal publishing focuses on synthesis and integration of the literature disclosed in piecemeal fashion in the preprints. The concept of the virtual journal mentioned above would take in the Model the form of "overlay" journals that link out to the preprint literature. See discussion elsewhere in this paper of how to understand "overlay" in the context of the Model.

The paper's later sections elaborate points in this section. For now, the purpose is to describe succinctly the overall vision of a possible publishing future and then, in the next section, address its rationale.

1. The ratio of journal articles to preprints greatly diminishes. Assuming that growth rates in physics preprints remain unabated, a significant contraction in the number of journals would accomplish this decrease in ratio. There will be a much higher bar in TJP for publication of new results.

Within its newly contracted space, TJP provides syntheses—that is, reviews--of new scientific developments initially disclosed in preprints. By contrast, preprints serve as the primary testing ground for new research still in a nascent stage. Preprints will supplant whatever role TJP has played in providing initial disclosure of results.

2. In the Model, TJP articles become literature reviews of two types. TJP Type 1 review articles consist of summaries of a particular research agenda of a specific author or a team of co-authors that has played itself out over a series of preprints. TJP Type 2 review articles, much more consistent with the usual concept of a review article in TJP, consist of high-level reviews of a swath of preprints that summarize the work of competing groups of research teams working on similar research agendas. Over and above highlighting new preprint literature, both types of review articles would of course also reference pertinent TJP literature.

A sub-species of Type 2 articles can focus on publishing reviews about inter-disciplinarity between various fields, thus handling the problem of disciplinary insularity.

3. While the primary purpose of preprints is to disclose parts of a research agenda, an increased number of preprints will be review articles that synthesize other preprints.[5] We can call these Preprint Type 1 and Type 2 articles, by analogy with the two types of TJP review articles mentioned earlier.

These two type of preprint review articles can reinforce the symbiosis between preprint and TJP formats by giving TJP editors content to recruit.

4. In the Model, what balance should there be between number of reviews and non-review articles? This depends on the need for review literature in physics. Bibliometric and survey work need to establish the appropriate intensity of publication of review literature and how many journals would really be necessary to achieve the goals of the Model.

Some quite crude searching in Web of Science (one source of bibliometric data) suggests that physicists do not rely on review articles nearly as much as biology, unless these are artifacts of how WOS classifies articles as being of the review type.

Here are the results of very crude topic searches in biology and physics in the Web of Science "SCI-EXPANDED" database for 1965-Present completed on March 31, 2019.

> # 4
> 58,328
> TOPIC: (biology)
> Refined by: DOCUMENT TYPES: ( REVIEW )
> Indexes=SCI-EXPANDED Timespan=All years

---

[5] This contradicts the policy currently at [bioRxiv](): "Can I post a review article on bioRxiv? bioRxiv is intended for rapid sharing of new research. Some review articles contain new data/analyses and may therefore be deemed appropriate. Reviews that solely summarize existing knowledge are not appropriate and neither are term papers, book excerpts, and undergraduate dissertations."

```
# 3
302,918
TOPIC: (biology)
Indexes=SCI-EXPANDED Timespan=All years
Edit

# 2
13,282
TOPIC: (physics)
Refined by: DOCUMENT TYPES: ( REVIEW )
Indexes=SCI-EXPANDED Timespan=All years

# 1
431,052
TOPIC: (physics)
Indexes=SCI-EXPANDED Timespan=All years
```

This raises an interesting question. Why is the percentage of physics related review articles so much lower than the same percentage for biology? Perhaps this reflects the presence of medically related articles. Or perhaps these disparities are an artifact of the way the document type "review" is assigned by the vendor. In any case, far more sophisticated use of Web of Science or other data than presented above would make for valuable bibliometric analysis in relation to the Model.

Second, despite this comparatively "low" number of review articles, *Reviews of Modern Physics* (RMP) [touts](#) itself as "the most highly cited Physical Review publication." This suggests that physicists regard review articles as having great value--even if the overall stock of review articles in physics (if the WOS results actually indicate anything) is not as great as in other fields.

5. Journals that observe "Ingelfinger's rule" do not accept manuscripts previously published in the form of preprints. The Model suggest observance of the *opposite* of this rule. TJP editors strongly encourage or better yet require prior publication in preprint format of research submitted to their journals for publication. Preprints then become a routinely a very important and systematic recruitment ground for TJP for editors seeing research to subject to traditional peer review. In recruiting content, editors make holistic assessments of data reflecting a variety of bibliometric variables, none of which singly provides a sufficient basis for recruiting a preprint for consideration for TJP. In line with the Model, the emphasis is on recruiting review articles about research that has appeared over a number of preprints.

6. Preprints accommodate the tendency of researchers to segment their research. For an overview of the salami-slicing phenomenon, see (Elsevier 2017), which cites (Office of Research Integrity).There may be an ambiguity in the use of the term salami-slicing, between its use to describe unethical repackaging of the same work, versus merely taking a research agenda and segmenting it in reports of its development. (Certainly, academic societies should play a role in sharpening the relevant distinctions in this context.) Where the line between ethical and unethical salami-slicing begins and ends is unclear. Below uses the term "segmenting" to connote acceptable (in the sense of ethically sound) publication of discrete parts of a research agenda, where each publication is a new disclosure of a substantive development.

Scientists want to stake a claim for successive versions of their research.[6] Preprints are an appropriate venue for segmented disclosure of results, but TJP will properly discourage the segmenting practice by requiring manuscript submissions to *synthesize* important research results that appeared in this sliced fashion across a variety of preprints. These synthesizing efforts will help in transmitting science to the next generation and keep a running, macro view of progress in sub-disciplines and research agenda. Strongly related to this goal, the review articles can also play a critical pedagogical one. Consider (Broad 1981, p. 1138, concerning a professor of genetics who is "current drawing up the outlines for a genetics course. 'Here the fragmentation is clearly unfortunate because students confronted with a half-dozen short papers have a hard time seeing the forest for the trees'."

Finally, researchers who are new to a field would do well to write review articles of the first kind, as a way to master the literature of their field. This latter activity is a very suitable contribution of graduate students who have to do literature reviews anyhow for their dissertations. A good literature review can serve as a valuable first publication. Also, review articles provide a way to help identify topics to research for grant applications.[7]

7.  Some persons might claim that the level of intensive peer review practiced by TJP for decades serves a useful purpose: it helps scientists monitor late-breaking developments in fields and it rejects low-quality research. Arguably, while that is true, a far better investment of time for that purpose would be in TJP peer review articles of the kinds mentioned that synthesize the preprint literature, or in writing such articles, or in writing preprint-published reviews of the scientific literature. TJP editors can recruit the latter for peer review for possible publication.

8. The Model retains important features of TJP but argues that the latter needs reconfiguring. TJP implicitly contains the following sets of priorities:

- publication of huge numbers articles in TJP that reflect the segmenting phenomenon (mentioned above) outweighs the benefit of concentrating energies on TJP as the primary venue for retrospective syntheses of research directions;
- universities are well-served by spending huge amounts for subscriptions to a vast sea of journals, despite the ever-spiraling costs of education for hard-working parents;
- time devoted to peer review is worth more than time spent doing research, mentoring new students, or writing type 1 or 2 review TJP or preprint articles.

As in any market, stakeholders have to make hard decisions about how to balance conflicting priorities in use of resources, including time and money. The Model assumes that the priorities implicit in TJP need considerable rebalancing in light of the opportunity costs involved in perpetuating dysfunctionalities in TJP.

**An analogy: the National Archives**

---

[6] They also want to give evidence of their productivity for tenure and promotion purposes, but publication of results in journal format appears to be the definitive way to do this.
[7] Thanks to Phil Hewitt for suggesting this point by mentioning a research group that does this.

Anyone who has done research in the U.S.'s National Archives (NA) may find compelling this analogy for what the Model suggests.

The NA is a huge repository that includes a massive amount of material of all kinds relating to the operation of government departments. This material is of quite mixed quality in terms of its value in historical research. Anyone who does research there does well to "front end" their research prior to visiting the NA, given the serious time and logistical constraints involved in wading through such material. For a library guide based on this writer's experience in doing research about early economic development efforts, as well as WW2 predictions of the state of the post-war economy, see "Using the National Archives". One does well to look for government-published inventories of National Archives materials relating to one's subject of interest. Especially useful are guides that provide a narrative of the archival contents.

Like guides to what is in the National Archives, review articles provide guideposts or markers to what is truly valuable and deserves attention by time-starved researchers. In the Model, the actual archive consists of preprints. Indeed, perhaps the name of the major physics preprint server—"arXiv"—is no accident!

The analogy is of course a bit of a stretch in certain respects, but nonetheless it helps to concretize the point of the Model.

**How the Model addresses problematic aspects of journal publishing**

The Model addresses a number of problems that beset journal publishing. Again, the points here invoke examples from physics, but many of the same points should apply, even if not always straightforwardly, to other areas of STEM publishing.

1. **The publishing glut**

The sheer glut of STEM articles published year over year makes it increasingly difficult for each new generation of researchers to assimilate the most important research results of the prior generation. One way to measure of the growth in physics articles year over year is to consult the INSPEC database. (Milojevic 2015,Fig. 5) has a graph depicting the "annual publication volume" of physics, astronomy and biomedicine.

It is quite possible for a diligent scientific genius with huge amounts of free time as well as training in good online searching practice to assimilate and distill the accumulated literature and then advance it. That combination, however, is probably a rarity, thus driving an understandable fixation on minute areas of research.

Difficulty in doing exhaustive searches of the literature also naturally increases the chances of replicating research—or (more cynically) of researchers knowingly passing off long forgotten research as one's own. For a study of research replication, see (Simkin 2011). For some good reflections on the glut and its implications for science, and some suggestions about how to maneuver within it, see (Boon 2017).

Despite having been a longstanding problem[8], *that* the sheer glut of articles in TJP is so problematic does not attract the attention it deserves by persons interested in the reform of scholarly publishing. It does not figure in much (***any?***) of the discussions about OA, which has often (though not entirely!) crowded out discussion of other aspects of the publishing ecosystem. Moreover, it is not clear that persons concerned about journal pricing have focused as much as they should on the sheer quantity of articles published, given that this helps drive up prices.

In part, as (Boon 2017) notes, the glut in publishing likely reflects the afore-mentioned tendency to "salami slice" research into a large number of discrete publications, each chronicling one small stage in an evolving research program.[9] Regarding the proliferation of articles within physics publishing, here are some data points that help establish its scale.

The Journal Citation Reports, made available by Clarivate, provides impact factor and other data for journals in a variety of areas, including physics. Even though Journal Citation Reports is selective in its coverage (see the notes about Bradford's Law here), the number of physics journals it covers in various sub-disciplines elicits the question: are so many journals really needed, especially given that arXiv is readily available?

The history of *Physical Review* is a metaphor for growth in the physics literature. In reviewing Paul Hartman's *A Memoir on the Physical Review: A History of the First Hundred Years*, (Assmus 1997, p. 355) mentions:

> … he [Hartman] shows the Review to be a very concrete example of Derek Price's rule of exponential growth for the number of pages published by scientists: in 1954 the *Review* split into two parts,…; less than ten years later the journal was divided into five parts, and now, in addition, it issues reports, communications, comments, and addenda, as well as *Physical Review Letters*, added in 1958.

**The Model's call to contract the number of journal articles aligns with the longstanding mission of librarians to serve as conservators of the cumulative intellectual record throughout time**. For the reasons given, they should be advocates for a serious diminishment in the number of STEM articles published.

The fact that we now have a dual universe of preprints and journal articles in physics makes searching more difficult because of the large number of items in each domain. Is an earlier version of an article that was finally published in a journal good enough to cite, or does one have to locate the later published version if it is not in arXiv? Even if it is, but it is not the actual publisher version, can we be assured that it is identical to the latter?

---

[8] (Broad 1981) mentions "the emergence of the Least Publishable Unit (LPU), a term associated with the shrinking length of papers. LPU is a euphemism in some circles for the fragmentation of data. A researcher publishes four short papers rather than one long one. This fragmentation contributes to a host of problems, not the least being the sheer growth of the literature. One estimate holds that *Index Medicus* for 1985 will weigh more than 1 ton."

[9] Concerning the salami-slicing phenomenon, (Boon 2017) mentions that researchers who engage in this "divide papers into the least publishable unit in order to lengthen their publication list, increase the chances of being cited, and increase the opportunity to publish in journals with a high impact factor. This further contributes to the volume of papers published." Note the role that the impact factor, discussed below, plays.

To counter the glut, there should be very active efforts to contract the number of journals and to expand the role of preprints within scientific communication. The Model regards the primary role of journals as providing syntheses of research previously published in preprints, with preprints serving as the wild, open frontier—a marketplace of ideas--in which researchers disclose new findings.

A counterargument is that the Model paradoxically compounds this problem of searchability by accommodating segmenting of research in preprint publications. That is, even if journals diminish in number, preprints will continue to increase in number because of segmenting, and so difficulties associated with searching the scientific literature will not go away. The Model, however, handles this problem face on. Persons wanting to do literature reviews that discover important earlier results can confine themselves to search the much-diminished space of physics articles published in the TJP format. Again, in the Model TJP will increasingly emphasize synthesis of research results, with links to preprints, thereby enhancing their discoverability and tying together multiple preprints associated with a particular research program or sub-discipline.

An editorial from *Applied Physics Letters (APL)* (Collins 2015a; see follow up in Collins 2015b) provides an example of how it is possible to reorient journal editorial policies and emphases. This initiative shows that it is possible to raise standards of publication with the explicit goal of reducing the quantity of publications, in this case motivated by quality considerations. Collins notes that "our internal studies have shown fluctuations in the quality of our publications that we seek to minimize." As a remedy, Collins notes (p. 010401-1) that:

> Moving forward, APL will strive to ensure published manuscripts accomplish one or more of the following:
> - Report research that makes a substantial advance in applied physics and closely related disciplines.
> - Advance new or emerging fields that influence the direction of applied science.
> - Develop innovative technology using underlying physical principles.
> - Present scientific advances that cross multiple disciplines generating new avenues of science dialogue.
> - Take critical steps toward real world applications.

With the exception of the final editorial goal listed here, given its special relevance to applied physics areas, these sounds like useful guidelines strict observance of which will facilitate a contracted physics journal market, *mutatis mutandis*.

Interestingly, (Collins 2015 a) notes a "streamlined process for paper transfer from *APL* to *AIP Advances*". One wonders where there is here an opportunity to create an overlay review journal in accord with the Model, that is an overlay journal that reviews and discusses trends in applied physics, with links to preprints of articles that otherwise would be transferred in this way. This could make for an interesting experiment.

See (Mallapaty 2018), who notes:

> The number of primary research papers published by Applied Physics Letters (APL) has halved in the past few years. While the cut is exceptional, it concides with a slight curtailment in output among the 68 high-quality journals tracked by the Nature Index: global article counts have declined by almost 5% since 2014, from 58,000 to just over 55,000 in 2017.

The article goes on to suggest reasons for this decline.

## 2. Journal pricing

The glut of publishing creates demand for increasing numbers of journals, which in turn has buttressed high STEM journal pricing and historical year over year increases in journal pricing. Have preprints challenged this longstanding regime? It seems unlikely that a study of physics journals since the advent of electronic preprints would show any noticeable downward pressure either on the number of physics journals or on their prices.

First, some recent data about journal price increases and then some comments about the larger context of debates about journal pricing and the open access movement.

(Bosch 2018) provides two sources of data for physics journal and astronomy price increases. Here is the data, which should give a sense of the pricing of journals in physics and astronomy and (from Scopus) a sense of the number of journals, though this may underestimate the number. The Scopus journal numbers again elicit the question: do we really need this many journals?

**TABLE 3: COST HISTORY FOR ONLINE TITLES IN CLARIVATE ANALYTICS (FORMERLY ISI) INDEXES**

| SUBJECT | AVG # OF TITLES 2015-17 | AVGE COST PER TITLE 2016 | AVG COST PER TITLE 2017 | % OF CHANGE 2016-17 | AVG COST PER TITLE 2018 | % OF CHANGE 2017-18 |
|---|---|---|---|---|---|---|
| Astronomy | 11 | 1,957 | 2,026 | 4 | 2024 | 0 |
| Physics | 97 | 4,251 | 4,447 | 5 | 4,029 | -9 |

**TABLE 8: TITLES INDEXED IN SCOPUS COST HISTORY BY LIBRARY OF CONGRESS SUBJECT**

| SUBJECT | AVG # OF TITLES 2016-18 | % OF CHANGE 2016-18 | AVG COST PER TITLE 2016 | AVG COST PER TITLE 2017 | % OF CHANGE 2017 | AVG COST PER TITLE 2018 | % OF CHANGE 2018 |
|---|---|---|---|---|---|---|---|
| Astronomy | 57 | -4 | 1,876 | 2.044 | 8 | 2,221 | 9 |
| Physics | 396 | 4 | 3,171 | 3,306 | 4 | 3,454 | 4 |

The decline in pricing indicated in the Clarivate data looks interesting, but the data from Scopus, which covers a much larger number of journals, does not cohere with it. Of course, the methodology for arriving at the price values used here bears scrutiny. No time series analysis of previous years of coverage of this type of data will be attempted here, but is a worthwhile project.

One might argue that had preprints not existed, the number of physics journals would be significantly higher, but that is dealing with otiose hypotheticals.

At the outset of the movement to make journal articles openly accessible (OA), hopes abounded that this would through some mysterious mechanism challenge the TJP price regime. Up to now, it is has not been clear that the latter has happened. While OA is intrinsically laudable, concerted focus on it has to some extent deflected attention from the issue of pricing as a barrier to immediate access as well as the glut of publishing. That is, many discussions of open access have often at least implicitly conflated OA with the analytically distinct issue of inflationary increases in journal subscription. This is not, however, by no means to suggest that *all* OA advocates overlook pricing issues[10], just that much OA rhetoric does so. Nor is to say that OA advocacy may actually bring down prices in a very discernible way. The point is just that the mechanism by which this will happen has puzzlingly been overlooked.

For a long time it has been clear that inelasticities in toll-access journal markets would merely replicate themselves in author charge pricing. In the face of OA rhetoric, publishers protect their revenue streams by enabling a publishing model that shifts payments for journal access from toll access subscriptions to author payments for OA for individual articles, in some cases subsidized by libraries.

Returning now to preprint publishing, in arXiv we have a well-organized platform that makes OA a huge amount of scientific research, much of which journals eventually publish. It is natural to ask whether preprints will finally realize whatever promise its advocates initially thought OA would have to challenge the TJP pricing regime.

If the Model succeeds, perhaps it will exert a downward pressure on journal pricing. One can only speculate on the mechanisms by which this would occur. Perhaps contraction in the number of journals, reflecting as it would an assessment that the *raison d'etre* for TJP has eroded with an increasing valuation of preprint publishing, would drive down journal prices. There might, however, a corresponding increase in the prices of the journals as publishers seek to maintain their erstwhile revenue streams by increasing the prices of the remaining field of journals. How these opposing forces would play themselves out is anyone's guess. Thus the subjunctival tone of this paragraph.[11]

---

[10] For example, cf. SPARC's interest in challenging the Big Deals consisting of large aggregates of journals.

### 3. TJP's human resource cost to libraries

TJP imposes a huge human resource cost on libraries imposed. Year after year, in the face of flattened library budgets, the rising costs of journals, and the ever-expanding number of journal articles, librarians across the world spend countless hours evaluating journal use data and negotiating journal deals. To do so, they compile spreadsheets of use data to determine what subscriptions to maintain. They then engage in often protracted, enervating rounds of negotiations with journal publishers. All of this involves a tremendous opportunity cost in terms of more creative and useful contributions that librarians can make in serving their client populations. In the Model, a contraction of journals will greatly reduce these human resource costs.

### 4. The woes of peer review

The article glut discussed in the previous section creates a huge demand for peer review, which carries with it costs. The sheer volume of peer review means less time for doing research, improving teaching, mentoring the next generation of researchers, and writing of review articles of type 1 or 2. As (Velterop 2018, p. 2)[12] mentions, "It is not unambiguously clear that pre-publication peer review benefits scientific progress enough to justify its high cost, in terms of money (the cost of journals) and of time (the opportunity cost of time spent reviewing manuscripts, many of them more often than once)." In discussing "Comments and open peer review", (Pepe 2017, p. 6) mentions "the ever increasing number of publications and average number of authors per paper, which renders the current peer-review paradigm unsustainable."

A contraction in the journals market would significantly help to decrease the amount of time devoted to peer review of the traditional kind. This point is independent of any assessment about whether peer review as standardly practiced is, in itself, problematic. *The Model is agnostic about the best form of journal peer review but says that if there is peer review, it should be limited to review articles.*

Rather than spending time on peer review of segmented publications of pieces of a research agenda initially disclosed in peer reviewed journal articles, a far better use of time by researchers would be to cooperate in writing type 1 and type 2 review articles, whether for preprints or journals, or in peer review review articles for journals. Preprint review articles become content that journal editors can recruit for peer review.

---

[11] As an addendum to above, for recent views about OA journals, see (Banks 2018). "As Open Access Week kicks off, **Physics World** talks to editorial board members of IOP Publishing's open-access journals about their views on the future direction of open-access publishing."

[12] Velterop provides an overview of the issues, bibliography, and a typology of types of peer review.

While this is not the place for an extended discussion of the vexed practice peer review, about which there are acres of literature, it is worth making a number of brief suggestive points.

First, any analysis of the value of traditional forms of peer review should examine its history. (Baldwin 2018) provides an historical account of the emergence and solidification of the peer review concept, including an etymology of the phrase. Implicit in Baldwin's (p. 439) mention of the following is a question about whether peer review as currently practiced hasn't assumed ridiculously exaggerated powers: "…an evocative headline about a 2012 *Physics Letters B* paper: 'CERN's Higgs Boson Discovery Passes Peer Review, Becomes Actual Science.'" Baldwin's article concludes (p. 558) that "peer review's perceived failures may have their roots in the gap between modern expectations of refereeing and the more modest functions it was initially designed to fulfill." It is also fascinating that "it was not until the 1960s that all *Physical Review* papers were sent out for external referee opinions." (p. 542).

Second, if Ginsparg is right, it is worth asking whether lack of peer review in preprint publishing are at least somewhat overblown: (Lariviere 2014, p. 1159) mentions a comment by Ginsparg that there is a kind of self-policing that occurs when posting e-prints, since "fear of ruining one's professional reputation" if the material posted is not quality material. Perhaps there needs to be more emphasis within academic societies in defining formal codes of honorable conduct with respect to preprint submissions, including expectation of high standards in the initial version of a preprint, to supplement whatever informal codes of conduct already operate.[13]

Third, the "Plan U" website makes a good point; Plan U is an effort to make it mandatory for granted funded research to be posted as preprints: "Importantly, the availability and permanent online archiving of manuscripts before their evaluation would also provide an opportunity for innovation in how peer review is organized and performed, and how it might be tailored to the needs of particular disciplines and audiences." This point holds of course even if Plan U is not implemented. (Incidentally, one of the commentators in (Michael 2019) claims that Plan U was initiated by the "founders of bioRxiv".)

Fourth, (Wang 2019) exemplifies the type of creative thinking that should go into consideration of peer review for servers like arXiv. It includes a literature review of such efforts. The scheme that the authors propose challenges the concept of "Community Peer Review" and suggests "Self-Organizing Peer Review." The scheme bears scrutiny of how it develops.

Finally, the organization PRT Peer Review Transparency calls for labeling publications with graphics that denote different types of peer review. A future arXiv could find ways to implement this labeling. This would help in bibliometric analysis of how quality of preprints, suitably defined, correlates with various levels of peer review.

---

[13] An astronomer mentioned that occasionally preprints appear that disclose half-baked research, presumably to exploit the primacy of disclosure that preprints provide. This type of misuse of the preprint format is not quite at the level of falsification of data, but it is entirely unfair to other workers in one's field.

## 5. Slow appearance of scientific results

Preprints enable the rapid disclosure of scientific research. They correct TJPs delay of access to new research, which can have two sources. The first is the time it takes to accomplish peer review. We can call this a supply side slowness. The second is that post-publication, toll-access/subscription walls can delay access by persons wanting to see the finished product of research; call this slowness in fulfillment of user needs the demand side.

It is not clear if the time it takes to accomplish peer review is really a problem for physics, given the ready availability of arXiv.[14] As a remedy to delays on the demand side, ILL is available at least in the developed world, but one wonders how many papers not ordered via ILL because of the perceived nuisance of initiating a request may be quite relevant to one's research.

A useful-looking initiative that attempts to make peer review more rapid is "sci*rev*", according to which: "Valuable new knowledge lays untouched at reviewers' desks and editorial offices for extended periods of time. This is an unnecessary loss of time in the scientific process which otherwise has become much more efficient. sci*rev* helps speeding up this process by making it more transparent." On the sci*rev* website, one can search by scientific disciplines such as physics and come up with a list of the review times provided by journals in that field.

The Plan U website claims, without providing citations, that "the early availability of new research on preprint servers allows other researchers, where appropriate, to begin building on the results immediately; estimates suggest the aggregate time saving could advance the pace of scientific discovery fivefold in 10 years." This is an interesting point, though one naturally wonders if these estimates are correct.

## 6. Excessive specialization in the sciences

Concerning specialization in physics, recall the comments cited above from the former Editor of *Physical Review Letters* Jack Sandweiss.

The publishing glut may not be the only creator of demand for journals, thereby driving up prices in the economic market they exemplify, whether we characterize it as oligopoly or monopolistic, depending on how far down we go into the sub-disciplinary specificity. Specialization may be another factor. Greater emphasis in TJP on distillation and synthesis of research results initially disclosed in preprints (type 1 and 2 TJP review articles) can help counter whatever excessive specialization there is in physics by providing a high-level view of cross-fertilization of sub-disciplines. Recall that one species of review journal that the Model advocates is one that tracks inter-disciplinary research disclosed in preprints.

---

[14]Collins 2015b, p. 160401-1 mentions that "with an average time to publication a bit under 80 days (averaged over 2014), APL [*Applied Physics Letters*] has been one of the fastest journals into print."  Guedon (2001, p. 52) notes "journals are rather inadequate when it comes to communicating quickly and efficiently; they are much better at validating and evaluating the relative worth of scientific authors. They are adequate to preserve the memory of science over the long haul (several centuries in the best of circumstances)". This statement is consistent with the historical view about the value of journals mentioned above, with the qualifier that to achieve the goal of preserving memory, journals need to be more and more focused on providing reviews of type 1 and 2, in order to help researchers navigate the burgeoning preprint literature and to contract TJP significantly.

**Barriers to preprints as vehicles for challenging the received TJP regime**

The previous section addressed the potentials preprints have to address problems with TJP. There are, however, very considerable barriers to actualizing the potential preprints have to reform TJP in the ways suggested earlier.

1. **Commercial and society economic interests**

No one appears to have established that preprint publishing has contracted the number of physics journals or decreased the pricing of physics journals. One major reason for why these have not happened is the economic interests of commercial and society publishers.

Might the society publishers be willing to contract their stock of journals? They will be reluctant to do so, given the role journals play in funding their income streams. Perhaps societies can find alternative revenue streams. For example, by creating preprint servers that provide some value add in reply to paying a nominal amount to post a preprint.

If scholarly publishing is as problematic as the points above suggest, there is every reason for societies to participate in its reform since this aligns with their mission to represent and advance scholarship and research. Note that the Model's emphasis on review articles helps advance their goal of promoting education. This may involve a creative effort to find sources of funding other than publishing.

Along these lines, see (Michael 2019), from which: "…with the publication of Plan S and the debates around its potential unintended consequences, the sustainability of publishing revenue, the financial cornerstone of many societies, could experience increased pressure. So this month we asked…the Chefs: How can societies be sustainable with all of the current pressures on their established revenue sources?" Concerning Plan S, see (Else 2018).

The following statement by AIP's John Haynes (from Gantz 2019) suggests creative roles for society publishers in advancing the agenda proposed in this essay.

> I think when you start to create such a large corpus of content which is freely and openly available, you need to consider how can you start to add value and add service on top of that by using some of the new technologies; artificial intelligence, natural language processing, machine learning; and apply those to that corpus of content. How can you start to extract data or extract meaning? Paul Allen, who is the Microsoft co-founder, is mining some aspects of the computer science literature on arXiv in a service called Semantic Scholar which has the stated aim to "cut through the clutter". There are opportunities for publishers to really think about how to add value across the whole corpus of content using these data mining techniques. Another forward-looking effort is starting to develop services to make it easier for authors. "I can submit to a preprint server and submit to a journal, and I don't have to do it twice because it's a bloody nuisance and it takes up too much of my precious time."

Haynes's suggestions above gesture toward new sources of income that to some extent can pick up the slack in revenues that accompanies the contracted journals market that the Model envisions.

In addition to points Haynes makes, the following are possible roles for physics societies:

1. Play a large and useful role in developing new bibliometrics and practices for assessing the quality of both preprints and journals[15] and then help *operationalize* them. This would involve much needed operationalizing of the large quantity of bibliometric work published each year and specifying research agendas tied closely to such issues as t and p evaluations or grant funding.
2. Use their pool of in-house member talent to organize objective and honest reviews of the work of candidates for t and p, or reports about research impact for grant funding agencies.
3. Play a role in ensuring the quality of preprint publishing by developing a code of ethics that establishes best practices for submission and use of preprints. This can help reinforce the sort of self-policing by members of the quality of preprints. (In relation to this point, see the comment by Ginsparg quoted in the section about peer review.)

### 2. University expenditures for TJP

Universities sustain TJP by making large expenditures for subscriptions and by demanding that faculty publish in high-impact [IMPACT DEFINED HOW?] peer reviewed journals as a condition for awarding tenure. Universities are also beholden to granting bodies that require evidence of prior quality of research among grant applicants. The significant involvement of various universities in publishing journals has not addressed the problems mentioned earlier. Journal publishing by universities have not challenged the overall price regime.

To help promote the Model, universities should decrease the number of journals to which they subscribe. Some of the money saved in this way can go toward funding preprint operations. arXiv relies on funding from institutional members.

### 3. The tenure and promotion system

This is not the place to engage debates about the strengths and weaknesses of the current tenure and promotion (t and p) system. If the system remains in its current form, then to actualize the Model there will have to be an expanded acceptance of preprints as a legitimate publication format. This is because if the Model succeeds, it will be more difficult to publish in journals. The latter will be fewer in number and the focus of journals will have shifted to synthesis and review of research initially disclosed in preprints.

---

[15] Cf. in this context (Guedon 2001, p. 60): "…it is interesting to note that Ginsparg knew well what information could emerge from the use statistics of his server, but he refused to relevse them for ethical reasons and political prudence. If evaluation were ever to rely on his archives, it had better emerge as a conscious, collective move stemming from the whole community of scientists [suggesting a role for societies!], and not from the initiative of a single individual. Now, time has come to build the evaluation tools on its own two feet, without meddling constraints coming from print-related concerns. Libraries can help."

To make preprints more acceptable in t and p decisions, reliable bibliometrics for evaluating and demonstrating their impact on science will need to emerge. Like analogous metrics already used in evaluating journal articles, any set of metrics used in evaluating preprints will have strengths and weaknesses. A holistic use of an assortment of metrics is therefore critical. Other sources of evaluation need to be used, such as external evaluation of experts as to the quality of research, but this too can have limitations. No one criterion will independently suffice. An important consideration is not to let quantitative data be the primary driver of a t and p decision, since it is so much subject to interpretation, as external evaluations are.

It is important in this context to discuss the problematic role the [impact factor](impact factor) (IF) plays in perpetuating the current TJP ecosystem. The allure of impact factors is that they provide an easy way to satisfy the need to hierarchize research. This is not just for t and p purposes, but also for other purposes: for libraries to justify subscription expenditures on the basis that no library cannot subscribe to journals that faculties know to have a high IF; for publishers to consolidate their "brand"; for institutional benchmarking; and for grant applications that mention a research agenda's prior publishing history.

One major problem of IFs is that publication in a high impact journal does not guarantee the quality of individual articles published in a journal. (Bohannon 2016) should lead us to ask to what extent an impact factor value primarily results from contributions by a subset of the articles in a given journal, making the IF a poor indicator of article quality, wholly aside from any questions it might pose about its accuracy for assessing journal quality. Why then should the IF of a journal in itself count in reviews of research quality for institutional benchmarking, if it is not known if the articles published in that journal actually contributed to the IF's value?

Elevating the importance of preprints, as the Model promotes, can indirectly undermine the sway of impact factors. To see why, consider first that preprints do not have the organization into journals that marks the TJP model. One might argue that classification of preprints by subject sets up *de facto* "journals" consisting of all preprints in a particular subject domain, but this differs from TJP's organization of articles within journals that have distinct submission and editorial policies and a volume and issue structure. All these journal characteristics encourage focus on the particular journal as a unit[16], therefore diverting attention from the quality of individual articles published within it.

An advantage of appropriately developed preprint bibliometrics is that they do not tie evaluation of research impact to the aggregate impact of a journal. Rather, they maintain focus on the quality of an individual article. For this reason, they can help subversively undermine the IF's undue sway through a feedback mechanism, since bibliometrics that focus on the quality of individual preprints can indirectly promote a focus on journal article metrics that shift attention away from aggregate journal metrics such as the IF to article-level metrics. In sum, to challenge the IF's sway, it is necessary to develop preprint bibliometrics to assess the quality of individual articles; more on this later.

---

[16] See (Guedon 2001, p.21) in this context.

The goal is not to create a Dadaistic publishing system in which "anything goes" in the sense of treating all articles as on par in quality. For those of us who are ontological realists in their philosophy of science, there is a recognition of a hierarchy of intrinsic research quality, based on the adequation of the research to reality. The Model merely shifts focus from the journal level to the article level.[17] The goal is greater accuracy and reliability in the bibliometric assessment of individual research contributions than the current system of IF-hierarchized journal titles affords. This goal extends to *both* sides of the Model: preprints as the place of initial idea disclosure, and journals as the place for synthesis and commentary on sub-disciplinary trends indicated in preprints and on research agendas spread across several preprints (type 1 and 2 review articles.)

Philosophically minded critics of the Model might argue that the Model's focus on individual articles is unduly nominalistic in its focus on the quality of individual articles. After all, they might argue, a journal represents a whole that transcends its parts, which are the articles that comprise it, and that this deconstructs a time-proven way of aggregating articles in journals, some at least of which have played long crucial historical roles in presenting physics research to the world. A plausible counterargument is, however, that the Model fully accommodates TJP, but newly reoriented to focus on providing integrative review, precisely to enable and actually advance that self-same historical role, lest it otherwise get lost in the great wash of journal articles. Once again, one rationale for the Model among others is to make it easier to transmit knowledge from one generation of researchers to another, precisely by perpetuating synthesis and review of the segmented findings of a research finding disclosed in preprints.[18]

Bibliometric assessment of preprints and journal articles should be symbiotic. For example, assessment of preprint quality could rely in part on favorable mentions of preprints within TJP review articles. There should be an algorithmic way to monitor such mentions. The section about enhancements to arXiv elaborates some concrete ways a preprint server can promote the collection of bibliometric data about preprints. Turning to review journal articles, it is important to develop a deeper bibliometric understanding of the role they play in physics.

(Mulligan 2013) provides a survey of attitudes about peer review, see), concluding (p. 149) that "while researchers recognize that peer review is imperfect, it appears that most believe it is the most effective mechanism for ensuring the reliability, integrity, and consistency of the scholarly literature."  One wonders whether support for peer review as currently practiced largely reflects habits of mind by persons acculturated into the t and p system, with its emphasis on publishing in high impact journals. Perhaps Ginsparg's point about the self-policing that occurs in preprint publishing (see section about peer review) already provides sufficient quality control for most purposes. To the extent it does not, would society sponsored codes of conduct for preprint publishing help enhance quality, exploiting as it would the ability of preprint servers to enable posting of successive versions of a paper as it improves in quality after feedback?

---

[17] (Guedon, 2001 p. 21) states that "the interest of commercial publishers is to keep pushing journal titles, and not individual articles, as they are the foundation for the financially lucrative technique of branding individual scientists."

[18] Philosophically, it is a long story and of course ranges beyond this essay, but there is a great need to recover the integrative, mereological understanding of part-whole relationships, and analysis and synthesis, that we find in Aristotle and Plato, on both epistemological and metaphysical planes. More practically: preprints=parts and analyses; TJP=wholes and syntheses.

These are open questions. It is important to note, however, that the Model does *not* rely on a critique of peer review.

## 4. Scientific A-historicity

Yet another major obstacle to implementing the Model will likely be the a-historical perspective of many scientists with respect to their sub-disciplines. This claim of course has very many exceptions. It is entirely understandable, however, that the drive to compete for grant funding and tenure and promotion creates a set of psychological dispositions immediately reactive to emerging cues in the research environment, namely, the latest literature in one's field and to the most recent trends or fads in grant funding. This is at the expense of a more integrative insight into the overall direction of research in one's sub-discipline whether in the last 3, 5, 10 or 20 years, or of its much deeper history in the twentieth century or even nineteenth century. Type 2 articles in the Model provide this insight.

A sobering corrective to a-historical tendencies is to consider Norton's (2012) depiction of Einstein's work. It is also worth recalling Maxwell's statement that "it is of great advantage to the student of any subject to read the original memoirs on that subject, for science is always most completely assimilated when it is in the nascent state, . . . ". (Maxwell 1954, p. xi).[19]

Type 1 review articles are more fine-grained than type 2 review articles, in that they focus on a particular research agenda pursued by a particular researcher or research team. Their role is to make it much easier to achieve an integrative understanding of the particular aspects of a research agenda disclosed in discrete segments in preprints. Reading a series of quality review articles published at various points in time can go a long way toward developing historical sensibilities about the progress of one's field or sub-field.

## 5. Editorial Policies

One barrier to reliance on preprints are those publishers that observe the Ingelfinger Rule, which disallows publication of a preprint prior to submission of a manuscript to a journal. It was pushback from various projects that quashed an early attempt to use preprints in biological or medical areas (see Cobb 2017.)

Major physics society publishers AIP and APS allow posting on an e-print server prior to publication.

---

[19] In this context, see History of Science in Learning and Teaching Science.

For physicists who want to identify publishers that still observe this rule, one starting point (but with a qualifier[20]) is Sherpa Romeo, which enables one to identify what physics journals allow posting of preprints. The website claims, "this listing characterises pre-prints as being the version of the paper before peer review and post-prints as being the version of the paper after peer-review, with revisions having been made."

**Enhancements to arXiv that can contribute to the Model's success**

The previous section discussed barriers to implementing the Model described above.

This section discusses enhancements to arXiv that can collectively help increase the prospect that preprints will assume the role that the Model ascribes them. Some of these enhancements will help promote the model by giving preprints more of the look and feel of journal articles. These enhancements will of course require considerable funding. As mentioned earlier, universities can take part of their savings from subscription reductions to help fund arXiv. Comparison of arXiv to other preprint servers, such as SSRN, can provide ideas for other enhancements not covered below.

Caveat: some of these features may already have been discussed or under development. The comments below relate to features that are readily or easily accessible for use at arXiv.

1. **arXiv overlay journals with corresponding email alert about new content**

The Model calls for continuing efforts to develop so called "overlay journals"; see (Marra 2017).

For the purposes here, overlay journals compile links to preprints that show promise of significantly influencing the future course of research. Overlay journals can have three flavors. One version consists of a collation of mere links to preprints that satisfy bibliometric criteria for quality or that a panel of editors handpick as being of interest. A second version provides annotations and a narrative that ties preprints together to create something approaching a type 2 review article (as discussed earlier) but not containing the same amount of detailed commentary. (An example from biology might be preLights.) A third version consists of full-fledged type 1 or type 2 review articles that are overlay journals insofar as they point to and comment on preprints in "meta" fashion.

2. **Data**

(Da Silva 2017) concludes that the "rush to publish work as a free OA document with a citable identifier, the DOI, may also invite a wealth of bad, weak, or poor science. To reduce this risk, given the centrality of preprints in the open science movement, preprints should also have open data policies, that is, preprints cannot be published unless the data sets are also placed in the public domain". This can be accomplished either by requiring that the data sets be available as ancillary files in arXiv or by requiring that preprint submissions include a data field, for empirically oriented articles. (Pepe 2017, p. 6) also asserts that "the arXiv of the future will host data and code alongside papers." Making data available in this way aligns with the ideals of the open data movement, including the interest in replicability of data. (Da Silva 2019, p. 162) mentions that "prperints have been marketed as a solution to the replication crisis…and thus serve as a replication-fixing tool by allowing biologists to challenge published results…, and prove their lack of reproducibility, or fortify and confirm their reproducibility, by presenting contrary or confirmatory data." Da Silva mentions biology here, and the issue of reproducibility in the social sciences is well known, but it is not clear whether the replicability issue has been fully addressed in physics and astronomy.

---

[20] When Sherpa Romeo, however, says that a publisher allows preprint posting, it is not clear (to this author) if this means (1) or (2):
(1) a preprint can be posted prior to (or concurrently with) submission of the article to a journal?
(2) a preprint version can be posted, but only *after* a journal has accepted or published the article? (where a preprint is an early version of a paper prior to peer review.)

If a preprint mentions data that resides elsewhere, arXiv could require a DOI for data sets that point to reliable data repositories, especially subject specific ones. This provides one more reasonable "bar" for submission that can help enhance the quality of submissions.

In a [webpage](...) about "ancillary" data, arXiv makes it clear that its overriding purpose is publishing articles and not data sets, even though to some extent it accommodates the latter.

There are plausible reasons for arXiv to encourage, even require (for empirical research) placement of the data within the repository and "alongside papers" (per Pepe):

- this makes it easier to ensure that data remains accessible, in the event that subject data repositories are no longer maintained; i.e., this promotes data preservation and stability of access.
- it might encourage standardization in how data is presented, by requiring use of a minimal set of metadata; the latter can be enforced by providing a form for uploading data, on which certain metadata fields are required.
- If it is true that making data available enhances citing numbers, then adding data might also enhance the citability of preprints. (For one discussion, see Piwowar 2013). Certainly this is a worthy way area for bibliometric study of whether the current data sets associated with papers in arXiv enhance citations.

For further observations about data, see (Pepe 2017, pp. 4-5). Pepe mentions:

> Data sharing has become a fundament practice across all scholarly disciplines. Simply, if a published research paper is built on data, the authors have to provide access to the minimal set of resources (data and code)….But sharing data in arXiv's "LaTeX to PDF" paradigm is not possible. A pilot to support data deposit alongside papers which was run at the arXiv from 2010 to 2013 (Mayernik et al., 2012), filed to gain traction. While the project had to face an unexpected cut in government support, we believe that part of its failure can be associated with the fact that the papers and the data were deposited as separate entities. How do people share data today? They use kludgy strategies. A growing trend in astronomy and physics, for example, is to link the dataset in the published or preprinted paper. This practice allows authors to make their data more visible and get credit for it, as it is linked inside the papers, but recent work shows that links rot quickly with time (Pepe et al., 2014).

See the point above about stability of access. (Mayernik 2012) mentions: " …, in a pilot illustration of external use of DC Instance APIs, the arXiv pre-print repository enables researchers to deposit data associated with articles. Upon deposit, the article remains with the arXiv system, while the data are deposited to the JHU Data Conservancy Instance. A bi-directional link is established between the paper in arXiv and the data in the JHU DC Instance. The arXiv pilot uses the search-and-access and ingest APIs."

3. **Bibliometric tools**

(Pepe 2017, p. 6) mentions that "**the arXiv of the future will be transparent and it will publish information about alternative metrics that may determine the true impact of a research paper** [emphasis theirs]."

Bibliometric data for preprints should have these purposes:

- enable greater reliance on preprints in tenure and promotion assessments
- help users of arXiv discover influential preprints
- make it easy for journal editors to recruit content from high impact preprints
- enable bibliometric identification of newly emerging disciplinary sub-domains in physics as well as other types of analyses that shed light on the practice of science and its coverage in databases.

arXiv already facilitates collection of bibliometric data that helps fulfill these purposes. Link-outs to Google Scholar and INSPIRE enable access to citing papers and determination of H indexes, and to data for cited and citing articles, respectively. Link-outs to ADS provide a number of metrics, including citing and cited data as well as co-reading data. Clicking on "metrics" in ADS brings up categories for "total", "normalized", "refereed", and "normalized refereed" citations. For t and p purposes, the number of times a "refereed" article cites a preprint can be valuable and even more valuable if the Model takes hold, since a preprint has to have significantly contributed to a sub-discipline to merit mention in a TJP review articles of the kind the Model advocates.

In implementing the Model, there is a need for bibliometric work about the role review articles play in physics and how they can be integrated with preprints into the physics publishing ecosystem. The analysis of citations to preprints exemplified in (Lariviere 2014) provides a framework for this type of research.

In addition to these existing capabilities, arXiv could include these bibliometric capabilities or features:

- consolidate citing data from ADS, Google Scholar, and INSPIRES to avoid overlap of data

- enable searching of cited references and their variants (to take into account typos); this would then allow identification of arXiv preprints that cite them,
- use this consolidated data to enable H index (Hirsch, 2005) calculations that incorporate both preprints and journal articles, with removal of preprints in this calculation that have been published; including both formats in an H index helps put the two formats on par in terms of their significance, reinforcing their symbiosis.
- provide altmetrics, by analogy with bioRxiv, according to which "altmetrics are provided that track attention to the article in blogs, tweets, news reports, and other media."
- sort articles by number of times cited
- measure a preprint's influence by taking into account its influence as measured not only by citing articles but also the types of metrics that go into Altmetrics (see e.g. Physical Review Letters for use of the latter).
- weigh a preprint's impact relative to a cluster of similar articles within the same classification code; cf. the discussion of "How to piggyback on the journal Impact Factor" in (Wang 2019, pp. 5-6).

Any of the bibliometric measures implicit in these suggested enhancements will have strengths and weaknesses. This is why holistic appraisal is crucial in any use of bibliometric data, as is expert judgment grounded in long experience.

### 4. Email alerts

Users of arXiv can already sign up for e-mailings of preprints that cluster around a particular topic. These, incidentally, constitute *de facto* "journals" for various sub-disciplines.

Periodic emailings of lists of preprints satisfying one or more relevant bibliometric filters (e.g., highly cited papers or ones that have typical characteristics of review articles) or that notify about new items that cite a specified preprint would help:

- editors recruit content for TJP
- scientists track citations to their own work or others' work
- journalists monitor late-breaking research in a given field.

A useful feature would be the ability to receive emailings of items that cite a particular work or author.

### 5. Bibliographic visualization

Providing the ability from with arXiv to visualize citing and cited relationships between citations, or to cluster them by concept, would enable tracking of progress in sub-disciplines. For examples of such visualization software, see (Simboli 2008) for mention of Eugene Garfield's "Histcite" and (Simboli  ) for a review of the now-defunct "RefViz" software, respectively. (Marra 2017, pp. 376-77) mentions the visualization website "PaperScape" (paperscapre.org) and the "wordclouds" website Cloudy Science (cloudscience.wordpress.com/).

Visualization of citing-cited relationships or of subject keyword clustering can help working scientists discover relevant preprints as well as monitor developments in their sub-disciplines. Also, it would help historians of science write narratives of the development of science, philosophers of science looking for case studies of how science progresses, and journalists who want to write well-informed popularizations of new trends to highlight in popular articles. So too can it help grant funders contextualize the research disclosed in preprints and mentioned in grant proposals, as well as tenure and promotion (t and p) committees and external parties asked to provide objective assessments (again for t and p purposes) of an individual's research.

### 6. Text mining

The arXiv documentation addresses issues about text mining, but it would be helpful to have a publicly accessible tool that enables the fascinating type of arXiv research that (Hand 2012) describes. arXiv could take a cue from JSTOR; concerning its tool to trace and visualize the history of a concept, (see King et al. 2012). (Pepe 2017, p. 4) mentions:

> While search engines are getting better at text mining PDFs, the chances that any current or future search engine will meaningfully extract and interpret text from a dense 2-column paper are low. Importantly, it is a futile exercise of reverse engineering. Why are we locking content in a format that is not machine-readable?

This issue is of course relevant to the extractability of large quantities of data in text mining.

### 7. Commenting and annotating features

For a discussion of "ArXiv-based commenting resources", see (Pepe 2017, p. 6.)  (Marra 2017) includes discussion of overlay journals. Marra (p. 384) comments:

> On the basis of the 2016 users survey, the ArXiv team appears now to be somehow mediating between researchers' 'conservative' and still prevailing attitude, focused on keeping the platform 'to the core mission', and an emerging 2.0 trend which favours innovations such as rating and commenting on top of it. The ArXiv-Next Generation initative, whose development has only just started…, might perhaps mark the beginning of a change in this respect, for as much as it's possible understand at present.

Unlike bioRxiv, which relies on the Disqus comments box software, ability to comment on arXiv articles appears to be indirect. Here is an [example](#) of an arXiv article that links out to the [Physics Overflow](#) service. Note the valuable blog linking capability exemplified in the [link](#) at this paper.

Comments boxes, and links to blog postings, can help authors improve their analyses or suggest counterarguments valuable in revising their work and subsequently posting a new version.  Per the Model, preprints that generate substantive discussion might indicate preprints worthy of recruit by TJP for review articles.  On the other hand, how to police abuses of the commenting boxes (e.g., spamming or removal of vindictive comments) poses a technical difficulty.

## 8. Granular classification or thesaurus schemes/abstracting and indexing

arXiv employs these classification schemes ( "MSC-class: math archives only"  and "ACM-class: required for cs archives only" ), but for physics an enhancement to arXiv would be to provide granular physics subject category headings. Examples of such schemes are APS's thesaurus, the PhySH - Physics Subject Headings  (see Conover 2016).   INSPEC has a thesaurus as well; see sample record.  The INSPEC classification is different in being a classification scheme and not a thesaurus. Implementation of a granular classification scheme or thesaurus terms would be useful in arXiv for a number of reasons:

1. They can play an educative role in demonstrating how the various subject areas covered in arXiv are organized.
2. They can help create a structured presentation and organization of preprints.
3. It can enhance the searchability of preprints. To enable this, persons submitting manuscripts to arXiv could be required to submit one or more classification code or thesaurus term to their manuscripts. At least some journals in physics already require this; requiring use of granular codes or subject terms can in a small way help break down some of the perceived distinction between preprints and journals.
4. It can be the basis for email alerts based on subject headings much more granular than the current arXiv basis for these alerts.  Use of them can enhance  and contributes to goal of putting preprints on par with journal articles to some extent

One might argue that use of classificatory or thesaurus schemes enforces a deleterious "silo-ing" of knowledge by confining it to narrow spaces. This is not the case. While not a classification scheme as much as a thesaurus, one thinks of how the searchable MeSH header database in PubMed's implementation of Medline help one conceptualize the hierarchical interrelationships of various medical concepts. In the MeSH records, there is more than one tree for a given subject header, which can help researchers visualize interdisciplinary connections between various medical concepts. Hierarchical thesaurus and classificatory schemes help researchers think systematically about where their research stands within their sub-discipline and how their work may relate to other areas. Their use also aligns with the goal of librarianship to encourage the organization of knowledge with the goal of promoting discoverability.[21]

There are additional advantages to deploying a more granular classificatory or thesaurus scheme in arXiv. It can help abstracting and indexing (A and I) databases in their efforts to index preprint literature. Editors of A and I databases can ensure there is a balanced coverage of preprints from a variety of subject areas and help them monitor newly emerging vocabulary terms for use in their own thesaurus and classification schemes.

Details about what commercial databases index arXiv would be helpful.

(Lariviere 2014, p. 1159) mentions:

> In 1997, arXiv began collaborating with the Astrophysics Data System (ADS) and the ADS created an index for astrophysics e-prints, making them available through the ADS abstracts service. In 2002, abstracts of all arXiv categories were included (Henneken et al., 2007). arXiv also has a relationship with SPIRES, the first electronic catalogue of grey literature, focused on high-energy physics preprints (Gentil Beccot, Mele, &Brooks, 2009). SPIRES counts citations to and from preprints and directs physicists to arXiv (82% of clicks from SPIRES go to arXiv) (Gentil-Beccot, Mele, & Brooks, 2009). SPIRES is currently being replaced with INSPIRE, which was created to "provide an even more flexible and extensible system to allow publishers, repositories, and researchers themselves to contribute and share information" (Brooks, 2009, p. 91).

See mention above about ADS and INSPIRE.

### 9. Manuscript template and uniformity of manuscript style

A possible enhancement to arXiv's guidelines for formatting submitted papers is to make available manuscript templates of the kind that journals provide. For example, AIP provides this template for *Journal of Applied Physics*. Requiring standardized formatting can play a role in creating a standardized look and feel for preprints that will help promote the Model's attempt to elevate the perceived status of preprints as well as underscore their symbiosis with journal articles, keeping in mind the standardized formatting in journals published on the TJP model.

Cooperation with producers (e.g., Overleaf) in making it easy to provide standardization of formatting in the context of a tool that enables easy use of Latex. (Note, incidentally, the mention here of Overleaf use in relation to *Applied Physics Letters*.)

### 10. Labeling of review articles

An arXiv webpage mentions the following:

> We are also rebuilding our technical infrastructure to improve the user experience, the adaptability of the system, and the scalability of moderation. One consideration we have been discussing as part of the Next Generation arXiv is to label the type of content posted to arXiv. Rather than trying to fit research articles, PhD theses, conference proceedings, etc. into the same bucket, we could label them and then let readers choose what content they want to read. This may allow our future moderation policies to evolve to accept a broader array of content.

In addition to the suggested labeling of the formats mentioned, it would be of great value—in line with the Model—to label articles as review articles of type 1 or type 2. This would facilitate ease of searching for review articles and enable editors of journals to recruit review article content for peer review. This is something that could be required as part of submission to arXiv. Other format types could also be required, such as whether an item is a conference proceeding. In general, labeling of format types helps in bibliometric studies.[22]

---

[22] Along these very general lines, the bioRxiv advanced search enables one to specify the type of Article: "new", "contradictory", and "confirmatory".

The question though is how to implement this labeling. One possibility is to require submitters of papers to identify to label articles as review article, on a suitable definition of review article. arXiv could even make the distinction of type 1 and type 2 a requirement. Another possible way to identify review articles submitted to arXiv would be to ask arXiv moderators to make this determination, though this may create too much work for them.

Yet another possibility is to create an algorithmic procedure to identify preprints review articles of type 1 and 2 (see above), then label them as such in arXiv. No such algorithm will be perfect. It would rely on bibliometric characteristics of review articles to label preprints as "possible reviews". The algorithmic identification of articles as being review articles can be useful to moderators, should it be easy for them to do a quick check of the accuracy of the algorithmic identification.

National Library of Medicine clarified in an email how articles acquire the "publication type" review. In addition to publisher designations that an article is a review, "the [NLM] Data Assurance group and the indexer are also able to apply the PT. A select number of citation go through a partial automation process and sometimes the PT is assigned at that stage. From an indexer point of view, we generally go by the author's wording, if they say it is a review, we index it as such. If the authors do not specifically say it is a review but it clearly is looking at the methods, we also apply the PT."

Algorithmic extraction of the word "review" would help identify review articles in arXiv. Other markers may also be relevant; one possible marker of a preprint that adopts a review focus is the cognitive, lexical or semantic density (or complexity) of its title. One can safely assume that shorter titles with more generic titles have a greater likelihood of being review articles, though this criterion (nor any, independently of examining the article) will not be foolproof. Perhaps the algorithmic could rely on a formula with weights for these bibliometric variables.

### 11. Moderated submission review, slightly expanded

arXiv has [volunteers](#) who [moderate](#) submissions. A role for moderators could be to ensure that acceptable submissions provide whatever future granularized subject classification or thesaurus indexing; see above. They can also establish whether a preprint is a type 1 or 2 review article if algorithmic identification of review articles is not feasible, or quickly confirm that algorithmic identification is accurate.

### 12. Manuscript submission capability

By comparison with bioRxiv's [B2J](#), arXiv could identify journals that allow direct submission of preprints to journals. Of course, the Model sees a value in enabling the ingesting of review focused preprints into journals dedicated to providing integrative reviews.

### 13. An end to versioning

There may be value in preprint authors indicating that a particular version of their preprint represents the final version. Possible values of doing so include: PUT IN

**Relevance of the Model for non-physics subjects**

A number of preprint servers now share with arXiv the –xiv suffix. Do the points above apply to other STEM areas? The answer depends in part on detailed scrutiny of why the subject areas covered in arXiv were the predominant early adopters of preprints while other fields have resisted them and why some still do not use preprints. To borrow a phrase mentioned in (Delfanti 2015, p. 1), why do some disciplines have a "'preprint culture'" while others do not? (Delfanti, p. 3, asserts that "in particle physics, archives apparently meet a demand for internal cohesion— which proves the sense of belonging—and internal competition.") While there is speculation here and there about these differences, a systematic approach is the purview of qualitative research in the sociology of science or "science studies". Another area for sociological study is to assess differences in uptake of preprints among sub-fields of physics and what explains them.

If preprints use becomes as widespread in other fields as it is in physics, there may be value in consolidating preprint publishing, whether in the science or social sciences, under the umbrella of arXiv. This would help especially in discovery of interdisciplinary linkages and relieve the need to search multiple preprint servers. A counterargument is that there may be value in having independent subject silos of information for reasons of manageability of the archive, and perhaps even data security concerns.

**Preprints and Science Librarianship**

The discussion above has mentioned ways the Model aligns with the goals of the library profession. Examples of concrete actions that science librarians can take to promote the Model include the following:

- use the increasing reliance on preprints as an argument in negotiations aimed to lower prices and to reduce the stock of subscribed journals[23]
- advocate use of bibliometrics that assess the potential for preprints to impact the progress of science and advocate the use of bibliometrics for t and p and grant funding. (Cf. Guedon, 2001)
- assist writers of review articles in doing literature searches (cf. (Rehtlefsen 2014) for the medical field)
- incorporate within their research skills instruction coverage of how to search the preprint literature and how to cite preprints
- creation of library guides about preprint publishing
- contribute to development of preprint servers by providing user experience insights, including those related to ease of use
- help to quell any residual faculty and graduate students concerns about publication in preprints

---

[23]For some discussion, see comments here: Re. the SPARC Landscape Analysis 2019 and new directions for library negotiations.

- suggest that doctoral students consider posting sections of their dissertations in preprint format, as a way to elicit feedback for purposes of revision.

**Conclusion**

To summarize, the Model promises to exploit the increasing role of preprints in scholarly communication while retaining the best features of TJP, but with the latter considerably diminished and reconfigured to help scientists track developments initially disclosed in preprints.

The barriers are formidable, and thus the focus on the potentials preprint have if they are integrated into an even more symbiotic relationship to journal publishing.

If these barriers to the Model prove too difficult to surmount in the near run (much more likely than not), at least the preprint enhancements discussed above are worth considering for their potential to develop a closer symbiosis between preprints and TJP, one that will fully actualize the unique contributions and appropriate roles of each. These enhancements may assist in small, incremental ways to realize the goals of the Model, or some of them at least.

Both universities and societies should work together to achieve the goals of the Model.

**Acknowledgments**

The usual disclaimer applies that none of the individuals mentioned here necessarily affirm this paper's views. My colleague Phil Hewitt discussed the role of review articles in physics versus biology and mentioned a research group that writes review articles as a way of mastering the literature and discerning research areas for grant funding. Jean-Claude Guedon (U. Montreal) qualified my comment on a listserv that OA advocacy has overlooked pricing issues. A recent panel discussion at Lehigh University gave insight into researcher perceptions about preprints. The ArXiv staff responded to a query about the presence of data sets in their repository.